\documentclass[twocolumn]{aastex63}

\usepackage{lineno}
\usepackage{booktabs}
\usepackage{multirow}
\usepackage{graphicx}   
\usepackage{hyperref}   

\newcommand{\nn}{\nonumber}

\submitjournal{ApJ Letters}

\shortauthors{Roy Choudhury \& Okumura}

\begin{document}

\title{ 
 Updated Cosmological Constraints in Extended Parameter Space with Planck PR4, DESI Baryon Acoustic Oscillations, and Supernovae: Dynamical Dark Energy, Neutrino Masses, Lensing Anomaly, and the Hubble Tension}
\email{
sroy@asiaa.sinica.edu.tw,
tokumura@asiaa.sinica.edu.tw
}

\author{Shouvik Roy Choudhury}
\affiliation{Institute of Astronomy and Astrophysics, Academia Sinica, No. 1, Section 4, Roosevelt Road, Taipei 106216, Taiwan}

\author{Teppei Okumura}
\affiliation{Institute of Astronomy and Astrophysics, Academia Sinica, No. 1, Section 4, Roosevelt Road, Taipei 106216, Taiwan}
\affiliation{Kavli Institute for the Physics and Mathematics of the Universe (WPI), UTIAS, The University of Tokyo, Chiba 277-8583, Japan}

\begin{abstract}

We present updated constraints on cosmological parameters in a 12-parameter model, extending the standard six-parameter $\Lambda$CDM by including dynamical dark energy (DE: $w_0$, $w_a$), the sum of neutrino masses ($\sum m_{\nu}$), the effective number of non-photon radiation species ($N_{\rm eff}$), the lensing amplitude scaling ($A_{\rm lens}$), and the running of the scalar spectral index ($\alpha_s$). For CMB data, we use the Planck PR4 (2020) HiLLiPoP and LoLLiPoP likelihoods, Planck PR4+ACT DR6 lensing, and Planck 2018 low-$\ell$ TT likelihoods, along with DESI DR1 BAO and Pantheon+ and DESY5 uncalibrated type Ia Supernovae (SNe) likelihoods. Key findings are the following: i) Contrary to DESI results, CMB+BAO+Pantheon+ data include a cosmological constant within $2\sigma$, while CMB+BAO+DESY5 excludes it at over $2\sigma$, indicating the dynamical nature of dark energy is not yet robust. Potential systematics in the DESY5 sample may drive this exclusion. ii) Some data combinations show a $1\sigma$+ detection of non-zero $\sum m_{\nu}$, indicating possible future detection. We also provide a robust upper bound of $\sum m_{\nu} \lesssim 0.3$ eV (95\% confidence limit (C.L.)). iii) With CMB+BAO+SNe, $A_{\rm lens} = 1$ is included at $2\sigma$ (albeit not at $1\sigma$), indicating no significant lensing anomaly in this extended cosmology with Planck PR4 likelihoods. iv) The Hubble tension persists at $3.2$–$3.9\sigma$, suggesting these simple extensions do not resolve it. v) The $S_8$ tension with DES Year 3 weak lensing is reduced to $1.4\sigma$, likely due to additional parameters and the Planck PR4 likelihoods.
\end{abstract}

\keywords{
cosmology: observations, dynamical dark energy, neutrino masses, lensing anomaly, Hubble tension
}

\section{Introduction}
\label{sec:1}
 While the six-parameter $\Lambda$CDM model has been widely considered the standard model of cosmology, the recent cosmological constraints from the Dark Energy Spectroscopic Instrument (DESI) collaboration \citep{DESI:2024mwx} have brought into question whether dark energy can be described by a cosmological constant, i.e., $\Lambda$. One popular choice for an evolving or dynamical dark energy is the $w_0 w_a$CDM model, where the equation of state (EoS) of dark energy (DE) is described by the  Chevallier-Polarski-Linder (CPL) parametrization \citep{Chevallier:2000qy,Linder:2002et}, given by $w(z)\equiv w_{0}+w_{a} ~z/(1+z)$, where $z$ is the redshift. One of the main results in the DESI collaboration paper shows that in the $w_0 w_a$CDM model, the cosmological constant is rejected by 2.5$\sigma$ and 3.9$\sigma$ when tested against the Pantheon+ \citep{Brout:2022vxf} and the Dark Energy Survey Year 5 (DESY5) \citep{DES:2024tys} supernovae datasets respectively (when combined with DESI BAO and CMB data). The tremendous implication of the said results have led to a large number of subsequent studies concerning dark energy \citep[see e.g.,][]{Tada:2024znt,Berghaus:2024kra,Park:2024jns,Yin:2024hba,Shlivko:2024llw,Cortes:2024lgw,DESI:2024kob,Carloni:2024zpl,Croker:2024jfg,Mukherjee:2024ryz,Roy:2024kni,Wang:2024dka,Gialamas:2024lyw,Orchard:2024bve,Giare:2024gpk,Dinda:2024ktd,Jiang:2024xnu,Reboucas:2024smm,Bhattacharya:2024hep,Pang:2024qyh,Ramadan:2024kmn,Wolf:2024eph,Efstathiou:2024xcq}.

 However, the DESI BAO collaboration results are not uncontested. In \cite{Cortes:2024lgw}, the authors point out that while the fit to $w(z)$ shows a dynamical nature, at the redshift ranges which are most strongly constrained by the data the DE EoS remains close to the $\Lambda$CDM value of $w(z)=-1$, to a precision around $\pm 0.02$. In \cite{Efstathiou:2024xcq}, the author points out possible systematics present in the DESY5 supernovae (SNe) data which might be leading to the preference for the non-$\Lambda$CDM behaviour. In particular, the author finds that the SNe common to both the Pantheon+ and DESY5 samples have an offset of $~0.04$ mag. between low and high redshifts and correcting this offset leads to DESY5 dataset being in agreement with $\Lambda$CDM, similar to Pantheon+. It is important to note that the analyses in this said paper have been done with SNe data alone (i.e., not in combination with BAO and CMB).

Apart from dark energy, neutrinos are important constituents of the universe. While neutrinos are massless in the standard model of particle physics, terrestrial neutrino oscillations experiments have strongly confirmed the presence of three different neutrino masses, of which at least two must be non-zero for neutrino flavour oscillations to occur. Given the mass-squared splittings $\Delta m_{21}^2 \simeq 7.42 \times 10^{-5}$ eV$^2$ and $|\Delta m_{31}^2| \simeq 2.51 \times 10^{-3}$ eV$^2$, there are two possible neutrino mass-orderings: Normal Ordering ($\sum m_{\nu}>$  0.057 eV) and Inverted Ordering ($\sum m_{\nu}>$  0.096 eV) \citep{Esteban:2020cvm}. Apart from the evidence for dynamical dark energy, one of the main results of the DESI BAO collaboration paper \citep{DESI:2024mwx} was the strongest bound on the sum of neutrino masses parameter till then, $\sum m_{\nu} < 0.072$ eV (95\% C.L.) in the minimal $\Lambda$CDM+ $\sum m_{\nu}$ model with a $\sum m_{\nu}>0$ prior against CMB+BAO data. This bound is much stronger than the bound of $\sum m_{\nu} < 0.18$ eV  circa 2016 \citep{Giusarma:2016phn}, $\sum m_{\nu} < 0.12$ eV circa 2018 \citep{RoyChoudhury:2018gay,RoyChoudhury:2018vnm,RoyChoudhury:2019hls,Vagnozzi:2017ovm,Vagnozzi:2018jhn,Giusarma:2018jei,Tanseri:2022zfe}, or the bound of $\sum m_{\nu} < 0.09$ eV circa 2021 \citep{DiValentino:2021hoh}. While the DESI bound is not strong enough to rule out Inverted Ordering at more than 2$\sigma$, later studies have put forward even stronger constraints with additional background probes \citep{Wang:2024hen,Jiang:2024viw}, all using Planck PR3 based likelihoods for CMB anisotropies. However, some other studies have shown that the DESI collaboration bound weakens somewhat when the new Planck PR4 likelihoods are used instead of Planck PR3 (2018) ones \citep{Allali:2024aiv,Naredo-Tuero:2024sgf,Shao:2024mag}, primarily due to the absence of the lensing anomaly in Planck PR4 \citep{Tristram:2023haj}. One also needs to take into account the fact that in the DESI collaboration results \citep{DESI:2024mwx} the $\Lambda$CDM model is disfavoured compared to the $w_0 w_a$CDM model, which begs the question whether one should take the bounds in the $\Lambda$CDM+$\sum m_{\nu}$ model seriously. The dynamical dark energy parameters are correlated with $\sum m_{\nu}$ (especially $w_a$) through the ``geometric degeneracy" \citep{RoyChoudhury:2019hls} (see equations 3.2 and 3.3, and the related text in this cited paper). The strong degeneracy between the DE EoS and neutrino masses was first noticed in \cite{Hannestad:2005gj}. While it is known that the quintessence-like or non-phantom dark energy ($w(z)\geq -1$ at all redshifts) leads to stronger than $\Lambda$CDM constraints on $\sum m_{\nu}$ \citep{RoyChoudhury:2018vnm,Vagnozzi:2018jhn,RoyChoudhury:2019hls,Jiang:2024viw}, the $w_0 - w_a$ parameter-space preferred in the DESI collaboration paper in a $w_0 w_a$CDM model falls outside the non-phantom/quintessence region \citep{DESI:2024mwx, Cortes:2024lgw}, and the bounds obtained on the $\sum m_{\nu}$ are relaxed by more than a factor of 2 compared to the ones in the $\Lambda$CDM+$\sum m_{\nu}$ model.

After neutrinos become non-relativistic at late times (due to their small masses), they start contributing to the matter density of the universe. They suppress the small scale structure by avoiding clustering at small scales due to high thermal velocities. This in turn affects the lensing of the CMB photons, effect of which shows up in the small scales ($l \geq 200$) of the CMB power spectrum as an enhancement of the two-point correlation functions \citep{RoyChoudhury:2020das,Lesgourgues:2012uu}. In a particular model, the theoretical prediction for the gravitational potential (responsible for generating the weak lensing of the CMB) corresponds to $A_{\rm lens} = 1$, where $A_{\rm lens}$ is the scaling of the lensing amplitude. When $A_{\rm lens}$ is varied as a cosmological parameter, the weak lensing is decoupled from the primary anisotropies that produce it and then scaled by the value of $A_{\rm lens}$ \citep{Calabrese:2008rt}. Thus, $A_{\rm lens}$ serves as a consistency check parameter. If the data favours $A_{\rm lens}>1$ in a particular model, it indicates a preference for more smoothing of the acoustic peaks in the power spectra (typically due to lensing) than what is theoretically expected. The ``lensing anomaly" in the Planck PR3 (2018) high-$l$ data \citep{https://doi.org/10.26131/irsa559,Planck:2018nkj} is well-known \citep{Planck:2018vyg}, where one finds $A_{\rm lens} = 1.18 \pm 0.065$ (68\%) in the $\Lambda$CDM+$A_{\rm lens}$ model. This value of $A_{\rm lens}$ is more than 2$\sigma$ away from the expected $A_{\rm lens} = 1$. This lensing anomaly, however, has been shown to have reduced to less than 1$\sigma$ in the same $\Lambda$CDM+$A_{\rm lens}$ model with the Planck PR4 (2020) likelihoods \citep{Tristram:2023haj}. Since massive neutrino suppress the matter power spectrum in small scales, increasing $\sum m_{\nu}$ decreases weak lensing of CMB photons, and thus $\sum m_{\nu}$ and $A_{\rm lens}$ are strongly correlated. This provides a strong motivation to use Planck PR4 (2020) likelihoods instead of Planck PR3 (2018). 

$A_{\rm lens}$ is also degenerate with the curvature density parameter $\Omega_k$ \citep{DiValentino:2019qzk}, and the lensing anomaly is intimately related to the curvature tenison \citep{DiValentino:2019qzk,Handley:2019tkm}, where Planck PR3 (2018) likelihoods alone preferred a closed universe, and $\Omega_k=0$ was rejected at more than 2$\sigma$. However, the curvature tension goes away once other datasets like BAO and SNe are included in the analysis \citep{Planck:2018vyg}. This is in sharp contrast to the lensing anomaly, which persists at more than 2$\sigma$ even when other datasets are included with Planck PR3. Thus in this work, we only consider the lensing anomaly while keeping $\Omega_k$ fixed at zero.

In this paper, we also focus on the Hubble tension, which is currently the one of the most discussed topics in cosmology. The Cepheid-calibrated type Ia Supernovae (SNeIa) based distance-ladder measurement of $H_0$ by the SH0ES collaboration provides $H_0 = 73.04 \pm 1.04$ km/s/Mpc \citep{Riess:2021jrx} (local universe measurement). This result is discrepant at the level of 4.6$\sigma$ with the Planck PR4 (2020) measurement of $H_0 = 67.64 \pm 0.52$ km/s/Mpc \citep{Tristram:2023haj} in the $\Lambda$CDM model. This discrepancy is known as the ``Hubble tension." The dynamical dark energy parameters are correlated with $H_0$ via the geometric degeneracy \citep{RoyChoudhury:2019hls,Efstathiou:1998xx,Vagnozzi:2019ezj} and might largely loosen the bounds on $H_0$ from CMB and BAO data. However, tighter constraints are expected when the uncalibrated SNeIa data from Pantheon+ or DESY5 are included, as these datasets constrain the late-time background evolution well, thereby partially breaking the degeneracy. At the same time, Hubble tension might also be alleviated to a certain extent by introducing extra radiation species in the early universe by adding to the $N_{\rm eff}$ (the effective number of non-photon radiation species in the early universe), as this increases the Hubble expansion rate in the pre-recombination era. However, $N_{\rm eff}$ also leads to a phase-shift of the CMB power spectrum towards larger scales, and a damping of the power spectrum that increases as we go to smaller scales \citep{Baumann:2018muz}. To completely resolve the Hubble tension, a contribution of $\Delta N_{\rm eff} \simeq 1$ is needed, which is currently strongly disfavoured by Planck data \citep{Planck:2018vyg,Vagnozzi:2023nrq} (and this is true even in the presence of neutrino self-interactions which can partially counter the effects of $N_{\rm eff}$ \citep{RoyChoudhury:2020dmd,RoyChoudhury:2022rva,Bostan:2023ped}. 

The $S_8$ tension, which is a discrepancy in the measurement of the amplitude of the matter clustering between low redshift weak lensing surveys and the high redshift CMB experiments, is another puzzle in cosmology which has received a significant attention in the cosmology community. In the $\Lambda$CDM model, with Planck PR3 likelihoods, the $S_8$ tension stands at the level of $\sim 2.4\sigma$ with DES Year 3 analysis \citep{Planck:2018vyg,DES:2021wwk}. However, with the Planck PR4 likelihoods, this tension drops to 1.5$\sigma$ \citep{Tristram:2023haj}. However, again, there has been no assessment of this tension in a dynamical dark energy cosmology with Planck PR4 likelihoods. Other than weak lensing, the disagreement with reshift space distortion (RSD) measurements and Planck PR3 remains at a moderate 2.2$\sigma$ level \citep{Nunes:2021ipq}. 
 
Given the above issues, it is important to have a closer look into the DESI collaboration results in \cite{DESI:2024mwx}. In this study, one of our main aim is to check whether the preference for dynamical dark energy survives in a largely extended parameter space where we allow for variation in usual extensions to $\Lambda$CDM: sum of neutrino masses ($\sum m_{\nu}$) and effective number of non-photon radiation species ($N_{\rm eff}$), the scaling of the lensing amplitude ($A_{\rm lens}$), and the running of the scalar spectral index ($\alpha_s$). This takes the total number of varying parameters to 12, which is double of the number of parameters in the vanilla $\Lambda$CDM model. For previous studies in such largely extended parameter spaces, see \cite{DiValentino:2015ola,DiValentino:2016hlg,DiValentino:2017zyq,Poulin:2018zxs,RoyChoudhury:2018vnm,DiValentino:2019dzu}. Another major aim of this paper is to provide a robust bound on the $\sum m_{\nu}$ parameter which can be used by the cosmology and particle physics community, by using an extended parameter space incorporating dynamical dark energy and also the scaling of the lensing amplitude parameter, $A_{\rm lens}$, which is degenerate with $\sum m_{\nu}$ \citep{RoyChoudhury:2019hls}. Thus including $A_{\rm lens}$ makes the bound obtained on $\sum m_{\nu}$ in this study more robust. At the same time, dark energy itself modifies the gravitational potential in the very late times, thus affecting the lensing of CMB photons. Till now, there has been no study in literature which has varied the $A_{\rm lens}$ parameter in a dynamical dark energy cosmology with Planck PR4 (2020) likelihoods. This provides us another motivation to include $A_{\rm lens}$ in our analysis, to give a definitive answer regarding the lensing anomaly in a dynamical dark energy scenario. Also, we do not expect to completely resolve the Hubble tension with this simple extensions to $\Lambda$CDM cosmology studied in this paper. One of our main goals in this paper is, rather, to assess the level of this Hubble discrepancy in our 12 parameter extended cosmological model with the new datasets. We also aim to assess the level of discrepancy with the weak lensing measuremnt of the $S_8$ parameter in a dynamical dark energy cosmology. With the release of Planck Public Release 4 (PR4) likelihoods (2020): HiLLiPoP and LoLLiPoP \citep{Tristram:2023haj}, Planck PR4 lensing combined with ACT DR6 lensing likelihoods \citep{ACT:2023kun}, DESI DR1 BAO likelihoods \citep{DESI:2024mwx}, and the latest uncalibrated type Ia Supernovae likelihoods: Pantheon+ \citep{Brout:2022vxf} and DESY5 \citep{DES:2024tys}, we think it is timely to update the constraints in such an extended model. The results will be undoubtedly useful to the cosmology and particle physics community.

Our paper is structured as follows: in section~\ref{sec:2}, we describe our analysis methodology. In section~\ref{sec:3}, we discuss our results from the statistical analysis. We conclude in section~\ref{sec:4}. Constraints on the cosmological parameters are tabulated in table~\ref{table:2}.

\section{Analysis methodology} \label{sec:2}

We note the cosmological model, parameter sampling and plotting codes, and priors on parameters in section~\ref{sec:2.1}. In section~\ref{sec:2.2}, we discuss the cosmological datasets used in this paper.

\subsection{Cosmological model and parameter sampling}\label{sec:2.1}

Here is the parameter vector for this extended model with 12 parameters :
\begin{eqnarray}\label{eqn:3}
	&\theta \equiv \left[\omega_c, ~\omega_b, ~\Theta_s^{*},~\tau, ~n_s, ~{\rm{ln}}(10^{10} A_s), \right. \nn \\ 
	& \qquad \qquad \quad w_{0, \rm DE}, w_{a, \rm DE}, 	\left.N_{\textrm{eff}}, \sum m_{\nu}, \alpha_s, A_{\textrm{lens}} \right].
\end{eqnarray}

The first six parameters correspond to the $\Lambda$CDM model: the present-day cold dark matter energy density, $\omega_c \equiv \Omega_c h^2$; the present-day baryon energy density, $\omega_b \equiv \Omega_b h^2$; the reionization optical depth, $\tau$; the scalar spectral index, $n_s$, and the amplitude of the primordial scalar power spectrum, $A_s$ (both evaluated at the pivot scale $k_* = 0.05~ \text{Mpc}^{-1}$); and $\Theta_s^{*}$, which represents the ratio between the sound horizon and the angular diameter distance at the time of photon decoupling. 

Rest of the six parameters are the ones with which we extend the $\Lambda$CDM cosmology. For the CPL parametrization for dark energy EoS, in this paper, we use the notation ($w_{0,\rm DE}$, $w_{a,\rm DE}$) interchangeably with ($w_0$, $w_a$). The other parameters, as noted in the introduction section, are the effective number of non-photon radiation species ($N_{\rm eff}$), the sum of neutrino masses ($\sum m_{\nu}$), the running of the scalar spectral index ($\alpha_s$), and the scaling of the lensing amplitude ($A_{\rm lens}$). 

We note here that we are using the degenerate hierarchy of neutrino masses ($m_i = \sum m_{\nu}/3$ for $i = 1,2,3$, i.e., all three neutrino masses are equal), and we are using a prior $\sum m_{\nu} \geq 0$. This choice is suitable considering that cosmological data is only sensitive to the neutrino energy density and hence the total mass sum \citep{Lesgourgues:2012uu}, and even the near future cosmological data would not be able to detect the neutrino mass-splittings since they are very small \citep{Archidiacono:2020dvx}. Also, forecasts show that in the event of an actual detection of $\sum m_{\nu}$ the assumption of degenerate hierarchy (instead of the true hierarchy) leads only to a negligible bias \citep{Archidiacono:2020dvx}. There is also no conclusive evidence for a particular neutrino mass hierarchy even when cosmological results are combined with other terrestrial datasets from particle physics, like neutrino oscillations or beta decay \citep{Gariazzo:2022ahe}.

We also note here that since we are varying the running of the scalar spectral index ($\alpha_s \equiv dn_s/d\rm{ln}k$, where $k$ is the wave number), we are in turn assuming a standard running power law model for the primordial primordial power spectrum of scalar perturbations, given by,

\begin{equation}
	\textrm{ln}\mathcal{P}_s (k) = \mathrm{ln}~ A_s + (n_s -1)~ \mathrm{ln}\left(\frac{k}{k_*}\right)+\frac{\alpha_s}{2} \left[ \mathrm{ln}\left(\frac{k}{k_*}\right) \right]^2.
\end{equation}
A small value of log$_{10}  |\alpha_s| = -3.2$ is naturally expected from slow-roll inflationary models \citep{Garcia-Bellido:2014gna}. However, it can be larger in certain other inflationary scenarios (see e.g. \cite{Easther:2006tv,Kohri:2014jma,Chung:2003iu}).

\textbf{Parameter Sampling}: We use the cosmological inference code Cobaya \citep{Torrado:2020dgo,2019ascl.soft10019T} for all the Markov Chain Monte Carlo (MCMC) analyses in this paper. For theoretical cosmology calculations, we use the Boltzmann solver CAMB \citep{Lewis:1999bs,Howlett:2012mh}. When utilizing the combined Planck PR4+ACT DR6 lensing likelihood, we apply the higher precision settings recommended by ACT. We use the the Gelman and Rubin statistics \citep{doi:10.1080/10618600.1998.10474787} to estimate the convergence of chains. All our chains reached the convergence criterion of $R-1<0.01$. We use Getdist \citep{Lewis:2019xzd} to derive the parameter constraints and plot the figures presented in this paper. We employ broad flat priors on the cosmological parameters, as given in the table~\ref{table:1}. 

\begin{table}
	\begin{center}
		\begin{tabular}{c c}
			\hline
			Parameter                    & Prior\\
			\hline
			$\Omega_{\rm b} h^2$         & [0.005, 0.1]\\
			$\Omega_{\rm c} h^2$         & [0.001, 0.99]\\
			$\tau$                       & [0.01, 0.8]\\
			$n_s$                        & [0.8, 1.2]\\
			${\rm{ln}}(10^{10} A_s)$         & [1.61, 3.91]\\
			100~$\Theta_s^{*}$             & [0.5, 10]\\ 
			$w_{0,\rm DE}$                        & [-3, 1]\\
			$w_{a,\rm DE}$                        & [-2, 2]\\
			$N_{\rm eff}$                & [2, 5]\\ 
			$\sum m_\nu$ (eV)            & [0, 5]\\
			$\alpha_s$                    & [-0.1, 0.1]\\
			$A_{\textrm{lens}}$          & [0.1, 2]\\
			\hline
		\end{tabular}
	\end{center}
	\caption{\label{table:1} Flat priors on the main cosmological parameters constrained in this paper.}
\end{table}

\subsection{Datasets}\label{sec:2.2}
\textbf{CMB: Planck Public Release (PR) 4}: We use the latest large-scale (low-$l$) and small-scale (high-$l$) Cosmic Microwave Background (CMB) temperature and E-mode polarization power spectra measurements from the Planck satellite. For the high-$l$ ($30<l<2500$) TT, TE, EE data, we use the latest HiLLiPoP likelihoods, as decribed in \cite{Tristram:2023haj}. For the low-$l$ ($l<30$) EE spectra, we use the latest LoLLiPoP likelihoods, described in the same paper \citep{Tristram:2023haj}. Both of these are based on the Planck Public Release (PR) 4, which is the latest reprocessing of both LFI and HFI instruments' data using a new common pipeline, NPIPE, leading to slightly more data, lower noise, and better
consistency between frequency channels \citep{Planck:2020olo}.  For low-$l$ TT spectra, we use the Commander likelihood from Planck 2018 collaboration \citep{Planck:2018vyg}. We denote the combination of these likelihoods as \textbf{``Planck PR4."}

\textbf{CMB lensing: Planck PR4+ACT DR6}. CMB experiments also measure the  power spectrum of the gravitational lensing potential, $C_l^{\phi \phi}$, using the 4-point correlation functions. We make use of the latest NPIPE PR4 Planck CMB lensing
reconstruction \citep{Carron:2022eyg} and the Data Release 6 (DR6) of the Atacama Cosmology Telescope (ACT) \citep{ACT:2023kun,ACT:2023ubw}. We employ the higher precision settings recommended by the ACT collaboration \citep{ACT:2023kun}. For brevity, we use the notation \textbf{``lensing"} to denote this dataset combination.

\textbf{BAO: DESI Data Release (DR) 1}. We use the latest measurement of the Baryon Acoustic Oscillation (BAO) signal from the Data Release 1 of the Dark Energy Spectroscopic Instrument (DESI) collaboration \citep{DESI:2024mwx}, which comprises of data from the Bright Galaxy Sample (BGS, $0.1 < z < 0.4$), the Luminous Red Galaxy Sample (LRG, $0.4 < z < 0.6$ and $0.6 < z < 0.8$), the Emission Line Galaxy Sample (ELG, $1.1 < z < 1.6$), the combined LRG and ELG Sample in a common redshift range (LRG+ELG, $0.8 < z < 1.1$), the Quasar Sample (QSO, $0.8 < z < 2.1$), and the Lyman-$\alpha$ Forest Sample (Ly$\alpha$, $1.77 < z < 4.16$). We denote this full dataset as \textbf{``DESI."}

\textbf{SNeIa: Pantheon+}. We use the most recent Supernovae Type-Ia (SNeIa) luminosity distance measurements from the Pantheon+ Sample \citep{Scolnic:2021amr}, which comprises of 1550 spectroscopically-confirmed SNeIa in the redshift range $0.001 < z < 2.26$. We utilize the public likelihood from \citep{Brout:2022vxf}. This likelihood includes the complete statistical and systematic covariance, imposing a constraint of $z>0.01$ to reduce the influence of peculiar velocities on the Hubble diagram. We denote this dataset as \textbf{``PAN+"}.

\textbf{SNeIa: DES Year 5}. We utilize the luminosity distance measurements from the latest supernovae sample comprising of 1635 photometrically-classified SNeIa with redshifts $0.1<z<1.3$, as publicly released by the Dark Energy Survey (DES), as part of their Year 5 data release \citep{DES:2024tys}. We denote this dataset as \textbf{``DESY5"}.

We note here that PAN+ and DESY5 have supernovae that are common to both samples. Thus the two datasets are never used together, to avoid any double counting.

\section{Numerical results} \label{sec:3}

The key results from our cosmological parameter estimation analyses are presented in table~\ref{table:2}, and visualised in figures~\ref{fig:1}-\ref{fig:10}. Here we describe our results regarding the cosmological parameters:

\begin{table*}
	
	\centering
	\resizebox{\textwidth}{!}{
		\begin{tabular}{ccccccc}
			\toprule
			\toprule
			\vspace{0.2cm}
			Parameter  &    Planck PR4  &    Planck PR4  &    Planck PR4  &    Planck PR4  &    Planck   \\\vspace{0.2cm}
			&     & + lensing &  + lensing +DESI &  +lensing  +DESI + PAN+  &    lensing + DESI + DESY5     \\
			\midrule
			\hspace{1mm}
			\vspace{ 0.2cm}
			$\Omega_b h^2$  &    $0.02227\pm0.00024$  &    $0.02235\pm0.00022$  &    $0.02237\pm0.00020$  &    $0.02242\pm0.00020$  &    $0.02239\pm0.00021$    \\ \vspace{ 0.2cm}
			
			$\Omega_c h^2$  &    $0.1184\pm0.0030$  &    $0.1193\pm0.0028$  &    $0.1192\pm0.0027$  &    $0.1191\pm0.0029$  &    $0.1190\pm0.0028$    \\ \vspace{ 0.2cm}
			
			$\tau$  &    $0.0579\pm0.0065$  &     $0.0582\pm0.0064$  &    $0.0583\pm0.0062$  &    $0.0585\pm0.0066$  &      $0.0584 \pm 0.0064$    \\ \vspace{ 0.2cm}
			
			$n_s$  &    $0.967\pm0.010$  &    $0.971\pm0.010$  &    $0.971\pm0.009$  &     $0.974\pm0.009$ &   $0.972\pm0.009$  \\ \vspace{ 0.2cm}
			
			${\rm{ln}}(10^{10} A_s)$  &    $3.038\pm0.017$  &    $3.043\pm0.015$  &    $3.044\pm0.016$  &   $3.043\pm0.016$   &     $3.042\pm0.016$    \\ \vspace{ 0.2cm}
			
			100~$\Theta_s^{*}$  &    $1.04079\pm0.00042$  &     $1.040758\pm0.00038$  &    $1.04079\pm0.00039$   &     $1.04081\pm0.00040$  &  $1.04082\pm0.00039$ \\ 

			\multirow{2}{*}{$\sum m_\nu$ (eV)}  &   \multirow{2}{*}{ $<0.700$ (2$\sigma$)}  &    $0.211^{+0.060}_{-0.21}$ (1$\sigma$),   &    $0.173^{+0.086}_{-0.12}$ (1$\sigma$),   &    \multirow{2}{*}{$<0.292$ (2$\sigma$) } &    $0.151^{+0.055}_{-0.14}$ (1$\sigma$),    \\ \vspace{ 0.2cm}
			&&  $<0.450$ (2$\sigma$) &  $<0.339$ (2$\sigma$) && $<0.318$ (2$\sigma$) \\ \vspace{0.2cm}
			
			$N_{\textrm{eff}}$  &    $3.03\pm0.21$  &    $3.09\pm 0.20$  &    $3.10\pm0.18$  &    $3.13\pm0.20$  &    $3.10\pm0.19$   \\ \vspace{ 0.2cm}
			
			$w_{0, \rm DE}$  &    $-01.09^{+0.67}_{-0.80}$  &    $-0.96\pm0.64$  &    $-0.45^{+0.32}_{-0.22}$  &    $-0.848\pm0.066$ &    $-0.745^{+0.070}_{-0.078}$   \\ \vspace{ 0.2cm}
			
			$w_{a, \rm DE}$  &    unconstrained  &    unconstrained  &    $<-0.37$ (2$\sigma$)  &    $-0.61^{+0.36}_{-0.29}$  &    $-0.95^{+0.41}_{-0.34}$    \\ \vspace{ 0.2cm}

			$\alpha_s$  &    $-0.0053\pm0.0078$  &     $-0.0031\pm0.0075$  &     $-0.0031\pm0.0072$ &   $-0.0022\pm0.0074$  &     $-0.0030\pm0.0074$   \\ \vspace{ 0.2cm}
			
			$A_{\textrm{lens}}$  &    $1.094^{+0.066}_{-0.11}$  &    $1.084^{+0.044}_{-0.090}$  &    $1.061^{+0.046}_{-0.052}$   &    $1.064^{+0.045}_{-0.053}$  &    $1.063^{+0.046}_{-0.053}$    \\ \vspace{ 0.2cm}

			$H_0$ (km/s/Mpc)  &    $71^{+10}_{-20}$  &    $73^{+10}_{-20}$  &    $64.6^{+2.3}_{-3.3}$  &     $68.1\pm1.1$    &    $67.2\pm1.1$   \\ \vspace{ 0.2cm}
			
			$S_8$  &    $0.774^{+0.051}_{-0.046}$  &    $0.787\pm0.040$  &    $0.822\pm0.022$  &    $0.809^{+0.020}_{-0.017}$  &    $0.811^{+0.022}_{-0.018}$    \\ 
			
			\bottomrule
			\bottomrule

	\end{tabular}}
	\caption{\label{table:2}\footnotesize Bounds on cosmological parameters in the 12 parameter extended model. Marginalised limits are given at 68\% C.L. whereas upper limits are given at 95\% C.L. Note that $H_0$ and $S_8$ are derived parameters.}
\end{table*}	

\subsection{$w_{0,\rm DE}$ and  $w_{a,\rm DE}$}
\begin{figure}[tbp]
	\centering 
	\includegraphics[width=.95\linewidth]{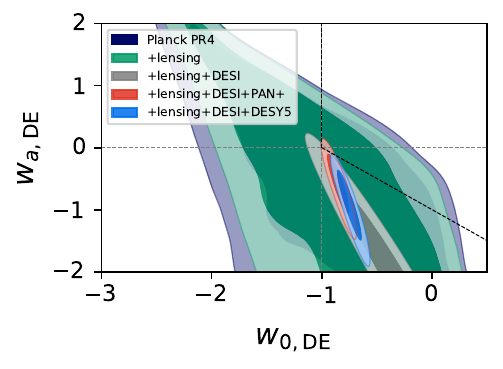}
	\caption{\label{fig:1} 68\% and 95\% marginalised  contours in the $w_{0,\rm DE} - w_{a,\rm DE}$ plane for different data combinations. The region at the bottom of the vertical dashed blue line and above the slanted dashed blue line is the parameter space for quintessence-like/non-phantom DE.}
\end{figure}

\begin{figure}[tbp]
	\centering 
	\includegraphics[width=.95\linewidth]{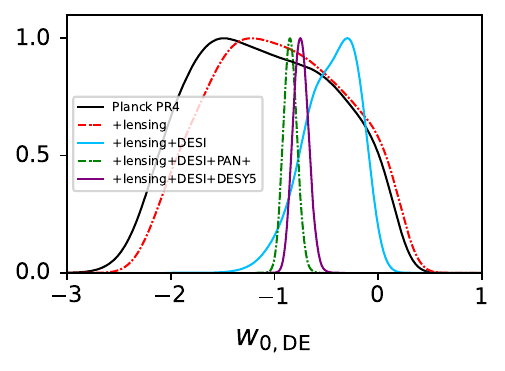}
	\hfill
	\includegraphics[width=.95\linewidth]{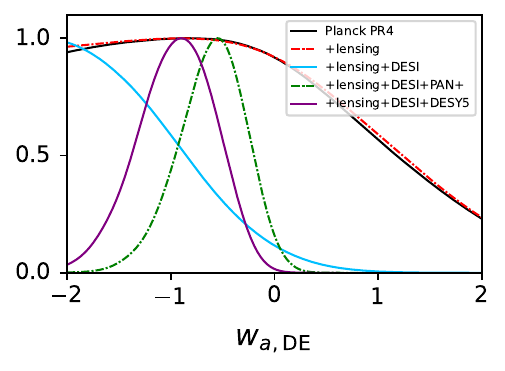}
	\caption{\label{fig:2}Comparison of 1-D marginalised posterior probability distributions for $w_0$ and $w_a$  for different data combinations.}
\end{figure}

As can be seen from figures~\ref{fig:1} and \ref{fig:2}, the dynamical DE EoS parameters remain poorly constrained without the use of SNeIa datasets. In fact, $w_a$ is unconstrained from both sides with CMB-only data. Very interestingly, we find that when we combine CMB and BAO data with Pantheon+, the cosmological constant ($w_{0,\rm DE} = -1, w_{a,\rm DE} = 0$) is allowed at 2$\sigma$. This is a significant result considering that the exclusion of the cosmological constant in the DESI BAO collaboration paper \citep{DESI:2024mwx} at more than 2$\sigma$ in a more restrictive 8-parameter $w_0w_a$CDM model. Our result implies that invoking simple extensions to $w_0wa$CDM cosmology can reconcile the cosmological constant again with the data. We hypothesize that there might be two distinct reasons for this: i) the degeneracy with the additional parameters might expand the allowed $w_0-w_a$ parameter space, ii) the Planck PR4 likelihoods might be partially responsible given that the DESI collaboration results are derived with Planck PR3. With the DES Y5 SNeIa data, we however find that the cosmological constant is rejected at more than 2$\sigma$, and the same is true for non-phantom/quintessence-like DE ($w(z)\geq -1$ at all redshifts). However, we advise caution, since, as we discussed in the introduction section, the DESY5 measurements might have unknown systematics as pointed out in \cite{Efstathiou:2024xcq}. Therefore, it can be concluded that the results from the DESI BAO collaboration regarding the evidence for dynamical nature of the dark energy EoS are not yet definitive.

\subsection{$\sum m_{\nu}$}
\begin{figure}[tbp]
	\centering 
	\includegraphics[width=.95\linewidth]{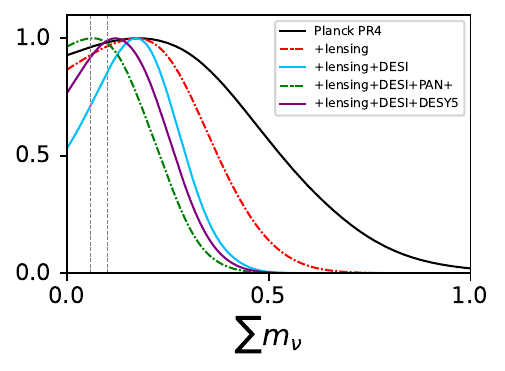}
	\caption{\label{fig:3}Comparison of 1-D marginalised posterior distributions for $\sum m_{\nu}$ [eV]  for different data combinations. The two vertical black dashed lines correspond to 0.057 eV and 0.096 eV, which are the minimum masses required by normal and inverted hierarchies respectively.}
\end{figure}

\begin{figure}[tbp]
	\centering 
	\includegraphics[width=.95\linewidth]{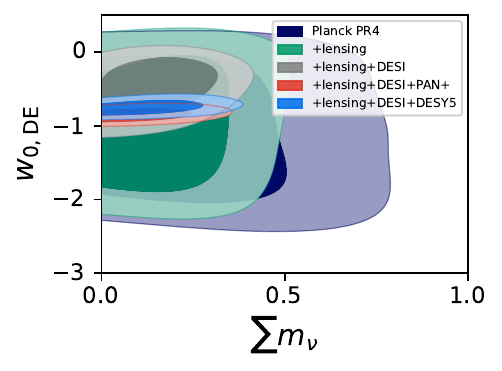}
	\hfill
	\includegraphics[width=.95\linewidth]{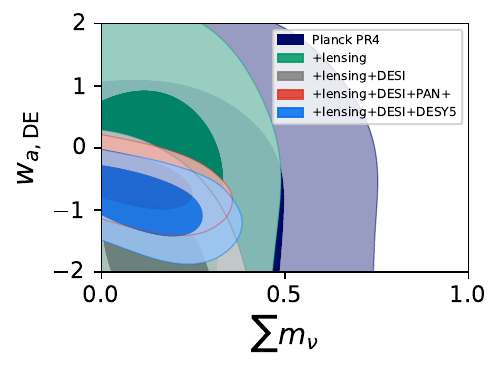}
	\caption{\label{fig:4} 68\% and 95\% marginalised  contours in the $w_{0,\rm DE}-\sum m_{\nu}$ [eV] (top) and $w_{a,\rm DE}-\sum m_{\nu}$ [eV] (bottom) planes for different data combinations.}
\end{figure}

The 1-D maginalized posterior distributions of $\sum m_{\nu}$ for various dataset combinations are given in figure~\ref{fig:3}, whereas in figure~\ref{fig:4} we present the 68\% and 95\% contour plots between $\sum m_{\nu}$ and the DE EoS parameters. While we do not have a 2$\sigma$ detection of non-zero neutrino masses, we find that each of the 1-D posterior distributions have a peak, and for 3 of the 5 dataset combinations studied in this paper, we get a 1$\sigma$ detection. This is encouraging, since it points to potential future detection of non-zero neutrino masses with cosmological data. At the same time, the bounds on $\sum m_{\nu}$ presented in this paper are much more relaxed compared to the bounds in the $\Lambda$CDM model in \cite{DESI:2024mwx}. While $w_{0,\rm DE}$ and $\sum m_{\nu}$ are only weakly correlated, $\sum m_{\nu}$ has a strong negative correlation with $w_{a,\rm DE}$. This is one of the primary reasons for the more relaxed bounds. There are other parameters which are degenerate with $\sum m_{\nu}$, as we show later. However, inclusion of those degenerate parameters also makes these bounds more robust, especially given the current uncertainty involving the nature of dark energy. With Planck PR4 data, the bound is quite relaxed at $\sum m_{\nu} < 0.700$ eV (95\%). This bound improves step-by-step, as we include lensing, BAO, and supernovae data. With Planck PR4+lensing+DESI+PAN+ we get a bound of $\sum m_{\nu} < 0.292$ eV (95\%), whereas with DESY5 instead of PAN+, the bound is very similar: $\sum m_{\nu} < 0.318$ eV (95\%). Given the robustness of these bounds, we suggest that the cosmology and particle physics community use this bound of $\sum m_{\nu}\lesssim 0.3$ eV (95\%) as a reference.

\subsection{$A_{\rm lens}$}
\begin{figure}[tbp]
	\centering 
	\includegraphics[width=.95\linewidth]{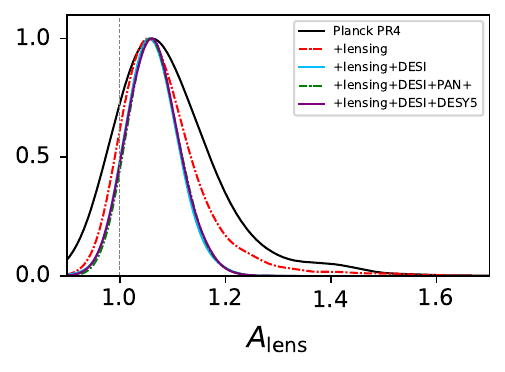}
	\hfill
	\includegraphics[width=.95\linewidth]{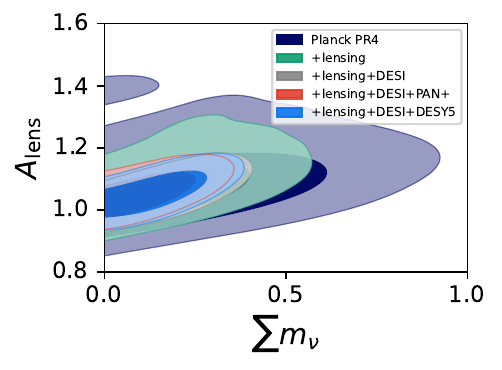}
	\caption{\label{fig:5} The top panel shows the 1-D marginalised posterior probability distributions for $A_{\rm lens}$ for different data combinations. The bottom panel shows the 68\% and 95\% marginalised  contours in the $A_{\rm lens}-\sum m_{\nu}$ [eV] plane for the same data combinations.}
\end{figure}

This measurement represents a crucial test of the consistency of the standard cosmological model, as deviations from  $A_{\rm lens} = 1$ could indicate new physics, such as changes in the growth of cosmic structures or modifications to gravity. The 1-D marginalised posterior distributions of $A_{\rm lens}$ and the correlation plots with $\sum m_{\nu}$ are given in figure~\ref{fig:5}. As expected, we find that $A_{\rm lens}$ has a strong positive correlation with the $\sum m_{\nu}$ parameter. More importantly, we find that all dataset combinations yield constraints on $A_{\rm lens}$ that are consistent with the theoretical expectation of $A_{\rm lens} = 1$ at less than 2$\sigma$. With Planck PR4+lensing+DESI+PAN+ we obtained a constraint of $A_{\rm lens} = 1.064^{+0.045}_{-0.053}$ (68\%). We get the almost same bound with DESY5 instead of PAN+. This highlights the reliability of our conclusions across different supernova samples. The absence of a significant deviation in the lensing amplitude implies that there is no strong evidence for a lensing anomaly within the context of a dynamical dark energy cosmology when using the latest Planck PR4 likelihoods. This is an important finding, especially when contrasted with earlier studies that employed Planck PR3 likelihoods, which had reported indications of a potential lensing anomaly.

\subsection{$N_{\rm eff}$}
\begin{figure}[tbp]
	\centering 
	\includegraphics[width=.95\linewidth]{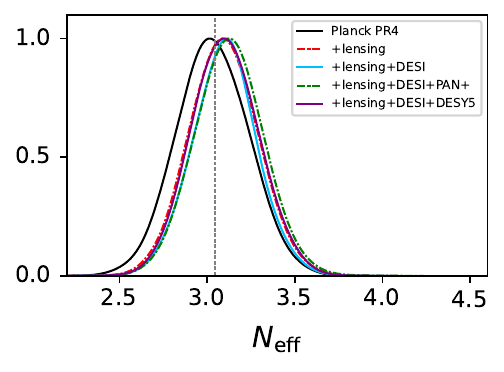}
	\hfill
	\includegraphics[width=.95\linewidth]{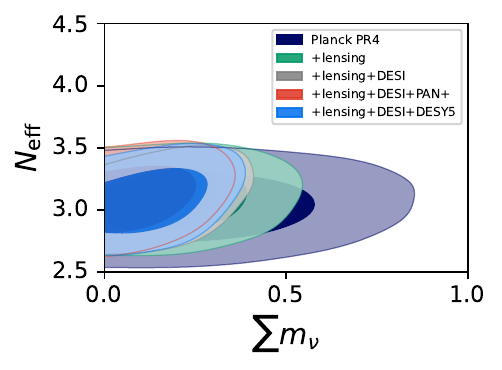}
	\caption{\label{fig:6} The top panel shows the 1-D marginalised posterior probability distributions for $N_{\rm eff}$ for different data combinations. The bottom panel shows the 68\% and 95\% marginalised  contours in the $N_{\rm eff}-\sum m_{\nu}$ [eV] plane for the same data combinations. The vertical dashed line in the top panel corresponds to $N_{\rm eff}^{\rm SM} = 3.044$. }
\end{figure}

The 1-D marginalised posterior distributions of $N_{\rm eff}$ and the correlation plots with $\sum m_{\nu}$ are given in figure~\ref{fig:6}. We find that all the obtained bounds on $N_{\rm eff}$ are fully consistent at $<1\sigma$ level with the theoretical value of $N_{\rm eff}$ assuming standard model of particle physics, i.e., $N_{\rm eff}^{\rm SM} = 3.044$ \citep{Froustey:2020mcq,Bennett:2020zkv,Akita:2020szl}. We also find that while $N_{\rm eff}$ does not have any significant correlation with $\sum m_{\nu}$, adding the SNeIa datasets leads to a slight tilt in the correlation direction in the 2D contour plots.

\subsection{$\alpha_s$}
\begin{figure}[tbp]
	\centering 
	\includegraphics[width=.95\linewidth]{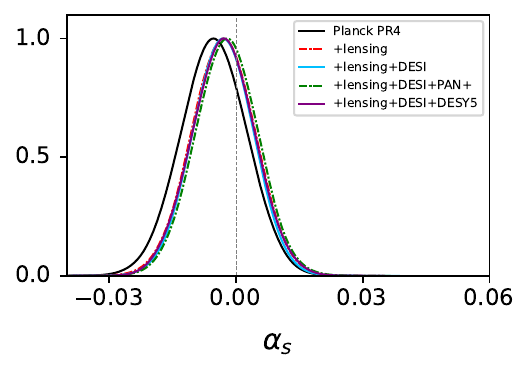}
	\caption{\label{fig:7}Comparison of 1-D marginalised posterior probability distributions for the running of the scalar spectral index $\alpha_s$ for different data combinations.}
\end{figure}

The 1-D marginalised posterior distributions of $\alpha_s$ for different data combinations are given in figure~\ref{fig:7}. We find that all the obtained bounds on $\alpha_s$ are consistent with $\alpha_s = 0$ within 1$\sigma$. This lack of significant running implies that more complex inflationary models with multiple fields or non-standard dynamics may not be necessary to explain the observed data. 

\subsection{$H_0$}
\begin{figure}[tbp]
	\centering 
	\includegraphics[width=.95\linewidth]{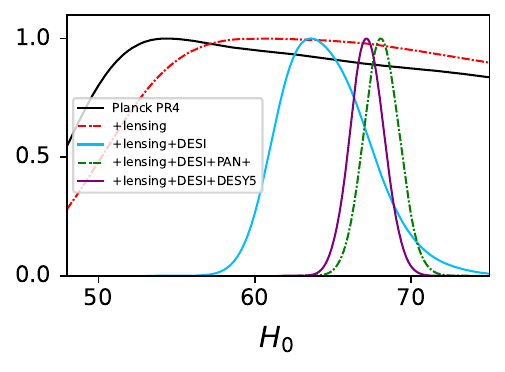}
	\caption{\label{fig:8}Comparison of 1-D marginalised posterior probability distributions for the the Hubble constant $H_0$ (km/s/Mpc) for different data combinations.}
\end{figure}

\begin{figure}[tbp]
	\centering 
	\includegraphics[width=.95\linewidth]{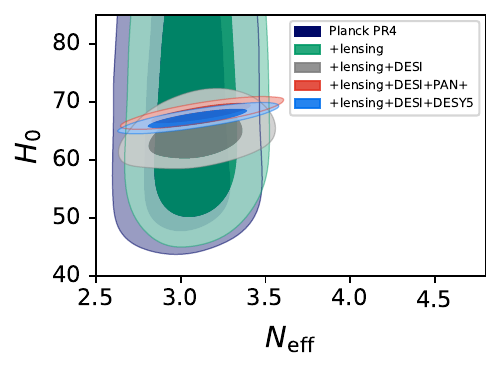}
	\hfill
	\includegraphics[width=.95\linewidth]{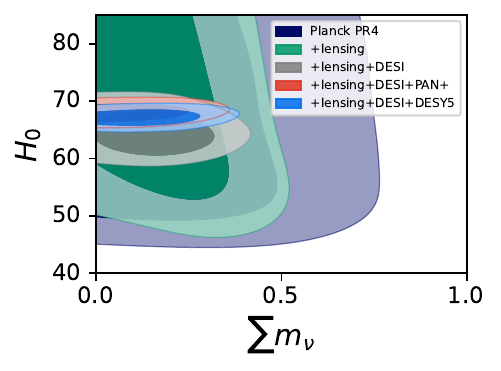}
	\caption{\label{fig:9} 68\% and 95\% marginalised  contours in the $H_0-N_{\rm eff}$ (top) and $H_0-\sum m_{\nu}$ [eV] (bottom) planes for different data combinations.}
\end{figure}

The 1-D marginalised posterior distributions of $H_0$ are given in figure~\ref{fig:8} and the 2-D correlation plots with $N_{\rm eff}$ and $\sum m_{\nu}$ are given in figure~\ref{fig:9}. We notice that CMB data alone cannot constrain $H_0$ and essentially a very large range of $H_0$ is allowed when only CMB data is used. The key reason for this is the strong degeneracy with the DE EoS parameters through the geometric degeneracy \citep{RoyChoudhury:2019hls,Efstathiou:1998xx}. However, we find that $H_0$ is also strongly correlated with $\sum m_{\nu}$ in the CMB-only data, through the same geometric degeneracy. But this degeneracy gets partially broken with DESI BAO data, and it is broken even further with the use of supernovae data, as can be clearly seen from the bottom panel of figure~\ref{fig:9}. This degeneracy-breaking quality of BAO and SNeIa have been studied in literature previously (see e.g. \cite{RoyChoudhury:2018gay}). On the other hand, CMB-only data produces no visible degeneracy in the $H_0 - N_{\rm eff}$ plot, since this degeneracy is weakened by the presence of the other degeneracies stated above. However, once the DE EoS parameter space is well-constrained with the use of BAO and SNeIa, we see the clear strong positive correlation between $H_0$ and $N_{\rm eff}$ in the top panel of figure~\ref{fig:9}. Regarding the Hubble tension, we note that there is a 3.2$\sigma$ tension between Planck PR4+lensing+DESI+PAN+ and the SH0ES \citep{Riess:2021jrx} measurement, which increases to 3.9$\sigma$ when DESY5 SNeIa are used instead of PAN+. Thus, it is clear that the simple extensions of $\Lambda$CDM studied in this paper are not sufficient to resolve the Hubble tension completely, i.e., below the 2$\sigma$ level. Therefore, the resolution of Hubble tension requires more complicated new physics (assuming data systematics are not responsible for the tension).

\subsection{$S_8$}
\begin{figure}[tbp]
	\centering 
	\includegraphics[width=.95\linewidth]{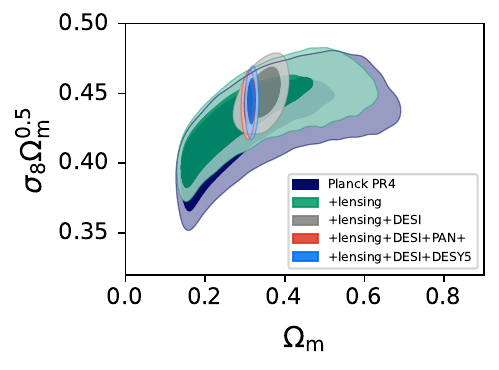}
	\caption{\label{fig:10}68\% and 95\% marginalised  contours in the $\Omega_m - \sqrt{0.3}S_8$ plane, where $S_8 \equiv \sigma_8(\Omega_m/0.3)^{0.5}$, for different data combinations.}
\end{figure}

Here, we concentrate on the tension associated with the amplitude of matter clustering in the late Universe, which is quantified by the parameter $S_8 \equiv \sigma_8\left(\frac{\Omega_m}{0.3}\right)^{0.5}$. In this expression, $\sigma_8$ represents the root mean square of the amplitude of matter perturbations smoothed over a scale of 8$h^{-1}$ Mpc, where $h$ is the Hubble constant in units of 100 km/s/Mpc, and $\Omega_m$ denotes the current matter density parameter. With Planck PR4+lensing+DESI+PAN+ we obtain a bound of $S_8 = 0.809^{+0.020}_{-0.017}$. This result is discrepant at the level of $\simeq$ 1.4$\sigma$ with the measurement derived from galaxy clustering and weak lensing from the Dark Energy Survey Year 3 analysis, $S_8 = 0.776\pm0.017$ (for $\Lambda$CDM with fixed $\sum m_{\nu}$) \citep{DES:2021wwk}. While the Planck PR3 dataset is in a $\sim 2.4\sigma$ tension with DES Year 3 results \citep{Planck:2018vyg, DES:2021wwk}, the reduction in this $S_8$ tension with Planck PR4 likelihoods is consistent with a previous result in literature \citep{Tristram:2023haj}, and the additional parameters considered in this paper might be increasing the error bar on $S_8$, thereby reducing the tension further, albeit slightly. In figure~\ref{fig:10}, we provide the 2D contour plots between $\Omega_m$ and $\sqrt{0.3}S_8 \equiv \sigma_8\Omega_m^{0.5}$.

\section{Conclusion} \label{sec:4}

We have presented updated constraints on cosmological parameters within a 12-parameter framework that extends the standard 6-parameter $\Lambda$CDM model to include additional degrees of freedom: the parameters for dynamical dark energy ($w_{0,\rm DE}$, $w_{a, DE}$), the sum of neutrino masses ($\sum m_{\nu}$), the effective number of non-photon radiation species ($N_{\rm eff}$), a scaling parameter for the lensing amplitude ($A_{\rm lens}$), and the running of the scalar spectral index ($\alpha_s$). For Cosmic Microwave Background (CMB) data, we utilized the latest Planck PR4 (2020) HiLLiPoP and LoLLiPoP likelihoods, in conjunction with Planck PR4+ACT DR6 lensing and Planck 2018 low-$\ell$ TT likelihoods. Additionally, we incorporated the latest DESI DR1 Baryon Acoustic Oscillation (BAO) likelihoods, as well as uncalibrated Type Ia Supernovae (SNe) data from the Pantheon+ and DES Year 5 samples.

Our primary findings are as follows:
\begin{itemize}
	\item  In contrast to the conclusions of the DESI collaboration, we found that the combination of CMB+BAO+Pantheon+ data does not exclude a cosmological constant at more than 2$\sigma$, whereas the combination of CMB+BAO+DESY5 data excludes it at a significance level greater than 2$\sigma$. Therefore, the evidence for a dynamical dark energy component is not yet compelling or robust. This discrepancy may be driven by potential systematics in the DESY5 SNe sample, as has been suggested in a recent study \citep{Efstathiou:2024xcq}.
	
	\item We observed that for certain data combinations, there is a 1$\sigma$ level indication for the sum of neutrino masses, $\sum m_{\nu}$, although this does not reach the 2$\sigma$ level detection threshold. Furthermore, all dataset combinations yielded a posterior peak, indicating a promising trend towards a possible future detection of non-zero neutrino masses. At the same time, we established a robust upper bound of $\sum m_{\nu} < 0.292$ eV at 95\% C.L. with CMB+BAO+Pantheon+, which is slightly more relaxed at $\sum m_{\nu} < 0.318$ eV with DESY5 SNe data. Given the current uncertainties regarding the true underlying cosmological model, we recommend that this bound of $\sum m_{\nu} \lesssim 0.3$ eV (95\%) is used by the cosmology and particle physics communities as a reference.
	
	\item Combining CMB, DESI BAO, and Pantheon+ data, we found that the scaling of the lensing amplitude parameter has the following constraint: $A_{\rm lens} = 1.064^{+0.045}_{-0.053}$ (68\%). With DESY5 SNe data instead of Pantheon+, we obtained a similar constraint. These results are consistent with the expected theoretical prediction of $A_{\rm lens} = 1$ at the 2$\sigma$ level (though not at 1$\sigma$). This implies that there is no significant lensing anomaly within the context of a dynamical dark energy cosmology when using Planck PR4 likelihoods (contrary to earlier findings in literature with Planck PR3 based likelihoods). The discrepancy between the PR3- and PR4-based results suggests that improvements in data quality and analysis techniques play a crucial role in resolving such anomalies. Therefore, our findings contribute to the ongoing efforts to refine our understanding of cosmic structure formation and the potential role of new physics in cosmology.

	\item Despite considering a significantly extended cosmological model, the Hubble tension persists at a significance level of 3.2$\sigma$ with CMB+BAO+Pantheon+ data combination, whereas the tension is at 3.9$\sigma$ with CMB+BAO+DESY5. Therefore, the simple extensions to $\Lambda$CDM explored in this study are insufficient to resolve the Hubble tension to below the 2$\sigma$ level.
	
	\item The $S_8$ tension with DES Year 3 analysis is reduced to $\simeq 1.4\sigma$ with CMB+BAO+SNe data. Reduction in tension to 1.5$\sigma$ from 2$\sigma$ level (with Planck PR3) has been previously documented with Planck PR4 in $\Lambda$CDM \citep{Tristram:2023haj}. Another reason for further reduction to 1.4$\sigma$ can be attributed to the inclusion of additional parameters (which increase the error bars on $S_8$) in our paper.
	
\end{itemize}

\acknowledgments
We acknowledge the use of the HPC facility at ASIAA (https://hpc.tiara.sinica.edu.tw/) where the numerical analyses were done.
TO acknowledges support from the Taiwan National Science and Technology Council under Grants No. NSTC 112-2112-M-001-034- and No. NSTC 113-2112-M-001-011-.

\bibliographystyle{apj}
\bibliography{biblio}




\end{document}